\documentclass{article}

\PassOptionsToPackage{numbers, compress}{natbib}

\usepackage[preprint]{neurips_2022}

\usepackage[utf8]{inputenc} 
\usepackage[T1]{fontenc}    
\usepackage{hyperref}       
\usepackage{url}            
\usepackage{booktabs}       
\usepackage{amsfonts}       
\usepackage{nicefrac}       
\usepackage{microtype}      
\usepackage{xcolor}         

\usepackage{graphicx} 
\usepackage{epstopdf} 
\usepackage{enumitem} 
\usepackage{amsmath,amssymb}

\title{Factor Investing with a Deep Multi-Factor Model}
\author{%
  Zikai Wei$^1$ 
  ~~~~
  Bo Dai$^2$
  ~~~~
  Dahua Lin$^{1,2}$\\
  $^1$The Chinese University of Hong Kong 
  ~~~~~
  $^2$Shanghai AI Laboratory\\
  \texttt{wz018@ie.cuhk.edu.hk}
  ~~~
  \texttt{daibo@pjlab.org.cn}
  ~~~
  \texttt{dhlin@ie.cuhk.edu.hk} \\
}

\begin{document}

\maketitle

\begin{abstract}
  
  Modeling and characterizing multiple factors is perhaps the most important step in achieving excess returns over market benchmarks. Both academia and industry are striving to find new factors that have good explanatory power for future stock returns and good stability of their predictive power. In practice, factor investing is still largely based on linear multi-factor models, although many deep learning methods show promising results compared to traditional methods in stock trend prediction and portfolio risk management. However, the existing non-linear methods have two drawbacks: 1) there is a lack of interpretation of the newly discovered factors, 2) the financial insights behind the mining process are unclear, making practitioners reluctant to apply the existing methods to factor investing. To address these two shortcomings, we develop a novel deep multi-factor model that adopts industry neutralization and market neutralization modules with clear financial insights, which help us easily build a dynamic and multi-relational stock graph in a hierarchical structure to learn the graph representation of stock relationships at different levels, \emph{e.g.}, industry level and universal level. Subsequently, graph attention modules are adopted to estimate a series of deep factors that maximize the cumulative factor returns. And a factor-attention module is developed to approximately compose the estimated deep factors from the input factors, as a way to interpret the deep factors explicitly. Extensive experiments on real-world stock market data demonstrate the effectiveness of our deep multi-factor model in the task of factor investing. 

\end{abstract}
\section{Introduction}

Factor investing is one of the most popular solutions in portfolio analysis and investment management for building an active portfolio with a desirable investment objective, 
which often relies on the linear multi-factor model as its core module \cite{levin1995stock,melas2018best,fama2020comparing}.
However, the assumption of a linear relationship between factor exposures and expected returns is quite restrictive, 
as it ignores interactions between different factors and does not always hold in reality,
and an increase in factor exposures may not lead to an increase in expected returns. 
Moreover, the linear multi-factor model characterizes the relationships of different stocks implicitly via their common factors, whereas their relationships are in fact much more complicated.

On the other hand, with the recent rapid development of deep neural networks, they have demonstrated great potential in learning the relationships between different factors \cite{wang2021review,vaswani2017attention}.
However, although deep neural networks have been widely applied to stock market prediction \cite{lin2021learning,chen2019investment,xu2021hist,xu2021rest,jiang2021applications},
there is only a factor investing method \cite{levin1995stock} dated back to 1995 that adopts neural networks to find the nonlinear relationship between factors and expected returns. 
One major reason that prevents practitioners in factoring investing from using deep neural networks is that the financial insights behind the deep models are not clear and often considered as a ``black box'', making the learned deep factors difficult to interpret. 

To introduce the advance of deep neural networks to factor investing without losing the underlying financial insights,
in this paper we propose a novel deep multi-factor framework that, through stock selection,
finds deep factors via graph attention modules on top of a multi-relational stock graph.
Specifically, the proposed framework consider stocks as nodes in a dynamic graph with two types of edges corresponding to stock relationships at different granularities.
In this way, the graph can account for the the unchronological changes in listing or de-listing of stocks and the changes in the multi-level relationships among stocks. 
The first type of edges connects nodes in the same industry to form an industry graph, from which we can learn the industry influence within the same industry. 
The second type of edges connects all nodes into a universe graph, from which we can learn the influence of other stocks with the same universe.
On top of this multi-relational stock graph, 
the proposed deep multi-factor model then adopts
graph attention module  summarize the industry and universe influences on other stocks.
Based on these influences, we can distill out the unique information from the stock context at different granularities, i.e., the original context, the unique context after neutralizing the industry influence, and the unique context after neutralizing the industry and universe influence.
Later, these contexts at different granularities can help the proposed model learn a set of deep factors to distinguish attractive and unattractive stocks and maximize the cumulative factor returns and information ratio.
Finally, 
a factor-attention module is further introduced to find the relationship between deep factors and the original input factors such as style factor, macroeconomic factors, 
which indicates how deep factors are composed from the original ones.
In this way, practitioners, such as fund managers, can  thoroughly understand the market logic and economic insights behind each deep factor used in stock selection.

To validate the effectiveness of our approach, we conduct a comprehensive study on real-world data containing three broad-based stock indexes which covers more than 2800 stocks in the Chinese stock market from 2010 to 2022. In summary, the contributions of this work include 1) developing a deep multi-factor model based on financial insights, where each component has a concrete financial meaning, 2) defining stock relationships via a dynamic and multi-relational graph with hierarchical levels of edges, 3) interpreting how our deep factor emerges from the original stock information.
\section{Related works}
The multifactor model has received considerable attention from researchers and academics for several decades to determine the exact nature of the common factors that influence risk and return in various assets and markets \cite{melas2018best}, where existing work classify it in two categories \emph{w.r.t} from different perspectives: linear vs nonlinear, and cross-sectional vs time-series. Comparing the cross-sectional and time-series models, it has been found that the cross-sectional factors can provide better explanation of the average stock returns than the model uses time-series factors \cite{fama2020comparing}. This inspires us develop a deep learning architecture to learn cross-sectional factors. 

With the rapid development of deep learning and its advantages in learning nonlinear relationships from big data in finance \cite{jiang2021applications, levin1995stock}, deep learning methods are applied to learn nonlinear relationships between style factors and expected returns of stocks \cite{nakagawa2018deep,nakagawa2019deep}. However, these methods do not consider the relationships between stocks in their models, and the financial insights behind their building blocks are not clear. To incorporate the multi-relational nature of stocks into our model, we turn our attention to graph neural networks \cite{velivckovic2017graph},
which have recently done great work in learning cross-stock relationships over a knowledge graph to solve a variety of finance problems \cite{wang2021review}, such as stock movement prediction \cite{xu2021hist,xu2021rest,lin2021learning,chen2019investment} event prediction, and risk management \cite{lin2021deep}. In our model, we use two graph attention networks to learn the influence of other stocks in the same industry and the influence of other stocks in the same universe.




\section{Method}

In this work, we develop a deep learning framework to find the cross-sectional factors that can consistently explain the average stock returns well, making good use of the relationships between stocks and the insights from finance.   

\subsection{Graph Construction} \label{graph_construction}
\emph{{Definition 1. Stock Graph.}} The stock graph is defined as a directed graph as in \cite{xu2021rest}, $\mathcal{G} = \langle \mathcal{S}, \mathcal{R},\mathcal{M} \rangle$, where $\mathcal{S}$ denotes the set of constituents of a broad-based stock index and $\mathcal{R}$ is the set of relations between two stocks. $\mathcal{M}$ is the set of adjacent matrices. For an adjacent matrix $\mathbf{M}^r \in \mathcal{M}$ of relation $r \in \mathcal{R}$, where $\mathbf{M}^r \in \mathbb{R}^{ | \mathcal{R}| \times | \mathcal{R}| }$, $\mathbf{M}^{r}_{ij}=1$ means that there is a relation $r$ from stock $s_j$ to stock $s_i$ and $\mathbf{M}^{r}_{ij}=0$ means that there is no such relation. However, the edges of the directed graph have an orientation \cite{wang2021review}, so this stock graph does not always hold in reality: stocks in the same industry always influence each other. In our work, we define the stock graph as a \emph{dynamic} and \emph{multi-relational} graph. The nodes and edges could be dynamic because the listed stocks change over time and a stock can be classified into a different industry, which leads to a change in the nodes and edges. We use a time-varying adjacent matrix $\mathbf{M}^r_t$ to represent this dynamic property.
We distill the stock context into different levels: Industry level and Universe level, where the edges at the different levels represent different types of relationships.
To learn the graph representation of stock relationships at different levels, we define the multi-related graph in a two-level hierarchy: an undirected graph describing the intra-industry relationship and an undirected graph interpreting the universe relationship.

\emph{{Definition 2. Industry Graph and Universe Graph.}} We define the industry graph as a stock graph $\mathcal{G}_s$ where stocks have relations to each other if they belong to the same industry. Similarly, the universe graph is defined as stock graph $\mathcal{G}_g$ that all constituents of a broad-based stock index are related. We use a time-variant graph for both cases of industry and universe, since the constituents in a industry or a index change over time.  

\subsection{Architecture}
\textbf{Stock Context Encoder.}
We define $n$ as the number of constituents inside a broad-based stock index, and $m$ as the number of types of stock context that covers fundamental, trading information and analysts' estimates. 
\begin{align}
    \mathbf{C}_t &= \text{MLP}\left(\text{BatchNorm}\left(\mathbf{F}_t\right)\right),
\end{align}
where $\mathbf{F}_t \in \mathbb{R}^{n \times m}$ is a matrix represents $m$ original factors \emph{w.r.t} $n$ stocks, and $\mathbf{C}_t \in \mathbb{R}^{n \times m_1}$ is the stock context matrix with $m_1$ hidden features extracted from the original factors. The financial insight of the batch normalization \cite{ioffe2015batch} used here is the similar to the \emph{z-score normalization} generally applied in data preprocessing for factor developing.

\textbf{Learning the Industry Influence.} Fundamentally, stocks are comparable inside the same industry, which makes the inner-industry relations important in industry-based analysis. Stocks in the same industry have mutual influence on each other beyond their common properties. To learn this industry influence, we apply Graph Attention Network (GAT) \cite{velivckovic2017graph} over the pre-defined industry graph, where the time-variant industry mask, $\mathbf{M}
_{t} \in \mathbb{R}^{n \times n }$, is symmetric and defined as
\begin{align}
\mathbf{M}_{ij}^{t} =
\begin{cases}
    1, & \text{if stock}\ s_{i} \text{ and stock} s_{j} \text{ are in the same industry}\\
    0, & \text{otherwise}
\end{cases},
\end{align}
where $t$ indicates the time stamp and $1 \le i,j \le n$. 
Then, we have the industry influence as
\begin{align}
    \mathbf{H}_{I}^{t} &= \text{GAT}(\mathbf{M}_t \mathbf{C}_{t} ).
\end{align}
This industry influence is useful for neutralizing the industry information.

\emph{{Definition 3. Industry Neutrality.}} 
We define the industry neutrality $\bar{\mathbf{C}}_{I}^{t}$ as the stock context $\mathbf{C}_t$ subtracted from the industry influence $\mathbf{H}_{I}^{t}$, \emph{i.e.},
\begin{align}
    \bar{\mathbf{C}}_{I}^{t} = \mathbf{C}_{t} -  \mathbf{H}_{I}^{t}.
\end{align}
Industry neutrality interprets the different context of a stock compared to other stocks in the same industry, which makes stocks comparable across different industries.

\textbf{Learning the Universe Influence.} Similar to learning the industry influence, we adopt GAT to capture the multual influence of any two stocks over the universe graph:
\begin{align}
    \mathbf{H}_{U}^{t} &= \text{GAT}(\bar{\mathbf{C}}_{I}^{t}),
\end{align}
where the universe is defined as a set of all constituents in a board-based stock index. The universe influence should learn from the stock context after eliminating industry influence, which can help to learn the cross-stock influence without industry effects.   

\emph{{Definition 4. Universe Neutrality.}} 
We define the universe neutrality $\bar{\mathbf{C}}_{U}^{t}$ as the industry-neutralized stock context $\mathbf{C}_t$ excluding the universe influence $\mathbf{H}_{U}^{t}$, 
\begin{align}
    \bar{\mathbf{C}}_{U}^{t} = \bar{\mathbf{C}}_{I}^{t} - \mathbf{H}_{U}^{t} .
\end{align}
The financial insight behind universal neutrality is that it contains the stock independent context with eliminating not only the effect of the industry on this stock but also the effect of the universe context.

\textbf{Learning Deep Factor from Hierarchical Contexts.} 
Formally, we learn the deep factor $\mathbf{f}_t \in \mathbb{R}^n$ from different granularity: stock original context $\mathbf{C}_{}^{t}$, industry neutralized context $\bar{\mathbf{C}}_{I}^{t}$, and universe-neutralized context $\bar{\mathbf{C}}_{U}^{t}$:
\begin{align}
\label{eq_factor}
    \mathbf{f}_{t} = \text{LeakyReLU} \left(\mathbf{W}^T \left(\ \mathbf{C}_{}^{t} || \bar{\mathbf{C}}_{I}^{t} || \bar{\mathbf{C}}_{U}^{t} \right)\right),
\end{align}
where $\mathbf{W}^T$ is a single-layer feed-forward neural network and $||$ represents concatenation.

\textbf{Multi-Head Learning.} We design $K$ output heads for learning ultimate factors corresponding to multiple forward periods be aware of practitioners, \emph{i.e.} $k$-forward trading days, where $k=3, 5, 10, 15, 20$. The Chinese mutual fund regulation does not allow fund managers sell the securities hold less than two trading dates. The minimum forward horizon, therefore, is set to three trading days. The multiple of five indicates the number of weeks that a security is hold. Based on Eq. \ref{eq_factor}, the deep factor learned \emph{w.r.t} different forward horizon is defined as
\begin{align}
    \mathbf{f}^{t}_{k} = \text{LeakyReLU} \left(\mathbf{W}_{k}^T \left(\ \mathbf{C}_{}^{t} || \bar{\mathbf{C}}_{I}^{t} || \bar{\mathbf{C}}_{U}^{t} \right)\right),
\end{align}
where $\mathbf{f}^{t}_{k} \in \mathbb{R}^n$ is the deep factor targets for explain the future information on the $k$-forward trading day and $\mathbf{W}^T_k$ is a single-layer feed-forward neural network corresponding to $k$-forward trading day.

\textbf{Interpretation of Deep Factor $\mathbf{f}^{t}$ via Factor Attention Module.} We introduce factor attention module to investigate the relationship between the deep factor and the original input factors or features. We employ attention mechanism \cite{vaswani2017attention} to attend our deep factor to learn the token importance of the original features. The normalized attention weight through a softmax function illustrates ``how much information influx in the deep factor from an original feature''.
\begin{align}
    \mathbf{U}^{t}_{k} &= \text{LeakyReLU} \left(\mathbf{W}_{a,k}^T  \mathbf{F}_t \right), \\
    \mathbf{A}^{t}_{k} &= \text{softmax} \left( \mathbf{U}^{t}_{k} \right), \\
    \bar{\mathbf{a}}^{t}_{k} &= \frac{1}{n} \sum_{i \in n} \mathbf{a}^{t}_{ik}
\end{align}
where $\mathbf{W}_{a,k}^T \in \mathbb{R}^{n \times n}$ is a single-layer feed-forward neural network corresponding to the $k$-forward trading day. $\mathbf{A}^{t}_{k} \in \mathbb{R}^{n \times m}$ is the attention weight matrix for the deep factor $\mathbf{f}^{t}_{k}$ and $\mathbf{a}^{t}_{ik}$ is $i$-th column of $\mathbf{A}^{t}_{k}$. Here, $\mathbf{a}^{t}_{k}$ is its corresponding attention weight vector of $m$. We define \emph{the estimated deep factor via factor attention}, $\hat{\mathbf{f}}^{t}_{k} \in \mathbb{R}^{m}$, as 
\begin{align}
    \hat{\mathbf{f}}^{t}_{k} =  \mathbf{F}_t^T \bar{\mathbf{a}}^{t}_{k},
\end{align}
where $\hat{\mathbf{f}}^{t}_{k}$ is the ``attention estimate'' of the deep factor $\mathbf{f}^{t}_{k}$. When $\hat{\mathbf{f}}^{t}_{k}$ is very close to $\mathbf{f}^{t}_{k}$, the attention weight $\bar{\mathbf{a}}^{t}_{k}$ can interpret the portion of quantity comes from the original input factors. 

\subsection{Loss Design}
Now we ponder the design of loss function to fulfill the learning objectives of deep factor.


\textbf {Maximizing Factor Stability.} One of our goals is to find a factor that has good stability of its predictive power, usually measured by the information ratio of information coefficient (ICIR) \cite{lin2021learning}.
We use $c_k$ to denote the ICIR of the deep factor $\mathbf{f}_k^t$, where $c_k$ measures the stability of the predictive power of the deep factor $\mathbf{f}_k^t$ for the future return over the next $k$ trading days $\textbf{r}_{t+k}$. Thus, since higher ICIR indicates higher stability, we look for a stable factor that maximizes ICIR. For more details on ICIR and information ratio, see our supplementary material.

\textbf{Maximizing Factor Return.} 
Factor returns are the cross-sectional regression coefficients \cite{lin2021deep}that indicate the return attributable to a particular common factor:
$\hat{\mathbf{r}}_{t+k} = {b^{t}_{k}} \mathbf{f}^{t}_{k} $,
where ${b}^{t} \in \mathbb{R}$ is the factor return at time $t$. We introduce a local optimizer to learn the parameter $b^{t}_{k}$, which is independent of the global optimizer. The deep factor is expected to have a higher cumulative factor return, $\sum_{t \in \mathcal{T}}b^{t}_{k}$.

\textbf{Improving ``Attention Estimate''.} We use $L_2$-norm to evaluate the deviation between the deep factor $\mathbf{f}^{t}_{k}$ and its corresponding attention estimate $ \hat{\mathbf{f}}^{t}_{k}$, which can be calculated as ${d}^{t}_{k} = {||\mathbf{f}^{t}_{k}- \hat{\mathbf{f}}^{t}_{k}||}_2$.

The overall loss function is 
\begin{align}
   \mathcal{L} = \frac{1}{|\mathcal{K}||\mathcal{T}| } \sum_{ k \in \mathcal{K} } \sum_{ t \in \mathcal{T} } {d}^{t}_{k} - {b}^{t}_{k} - {c}_{k} 
\end{align}
where $\mathcal{T}$ is the set of dates in the training period, $\mathcal{K}$ is the set of forward horizons.
 
\subsection{Optimization}
\textbf{Cross-sectional optimizer.} 
The cross-sectional optimizer is a local optimizer designed to learn a cross-sectional regression coefficient (factor returns) at each time $t \in \mathcal{T}$, by minimizing the mean squared errors of the predicted returns and the target returns. This cross-sectional optimizer is independent of the global optimizer we use to train our deep multifactor model.

\textbf{Global optimizer.} The global optimizer is designed to update the parameters of our deep multifactor model. It is set to simultaneously optimize the parameters in the stock context encoder, the industry and universe influences learning modules, the multi-head module, and the factor attention module.

\section{Experiment}

\begin{figure}[!htb]
   \centering
\begin{tabular}{ccc}
\includegraphics[width=0.31\textwidth]{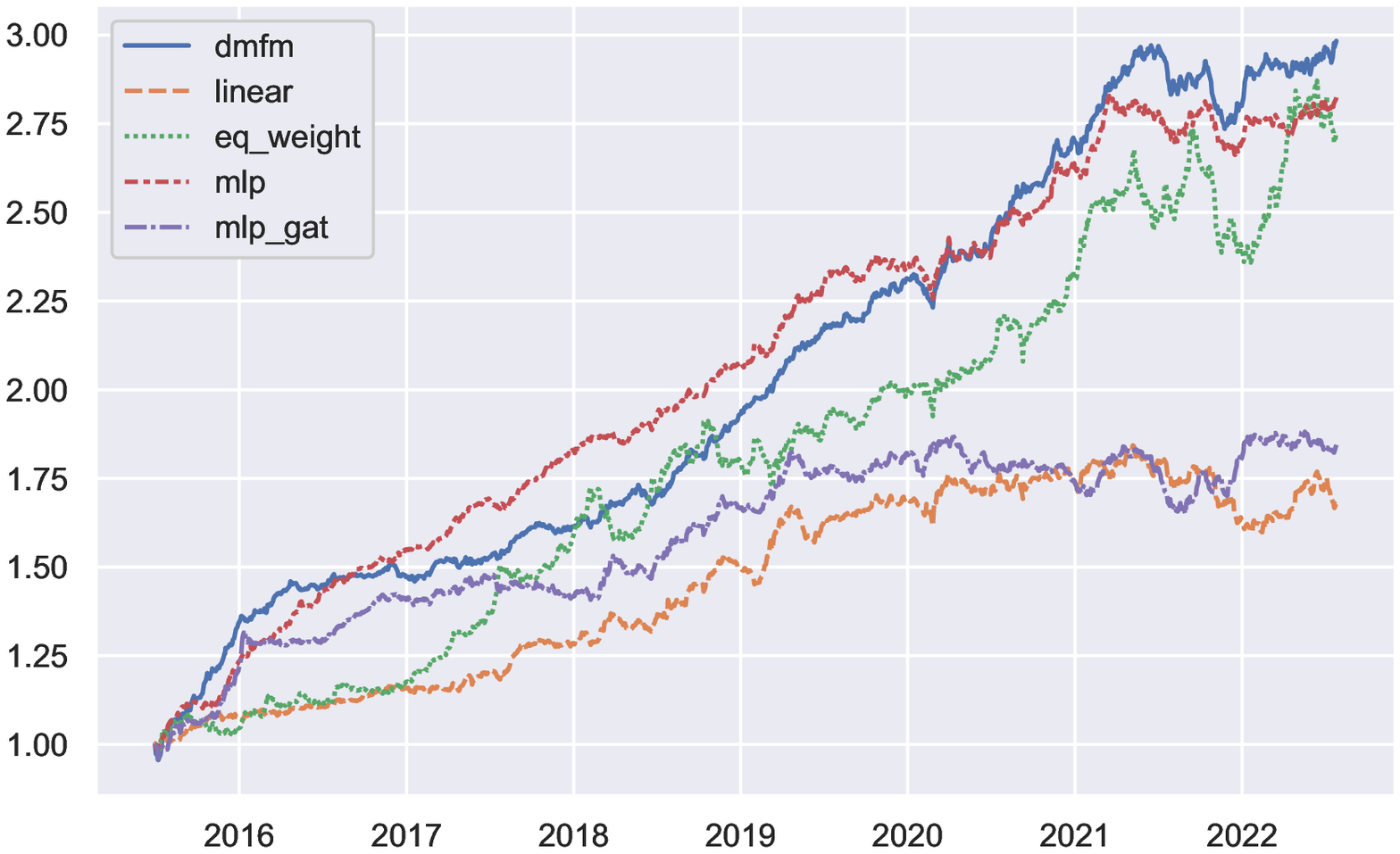}
&
\includegraphics[width=0.31\textwidth]{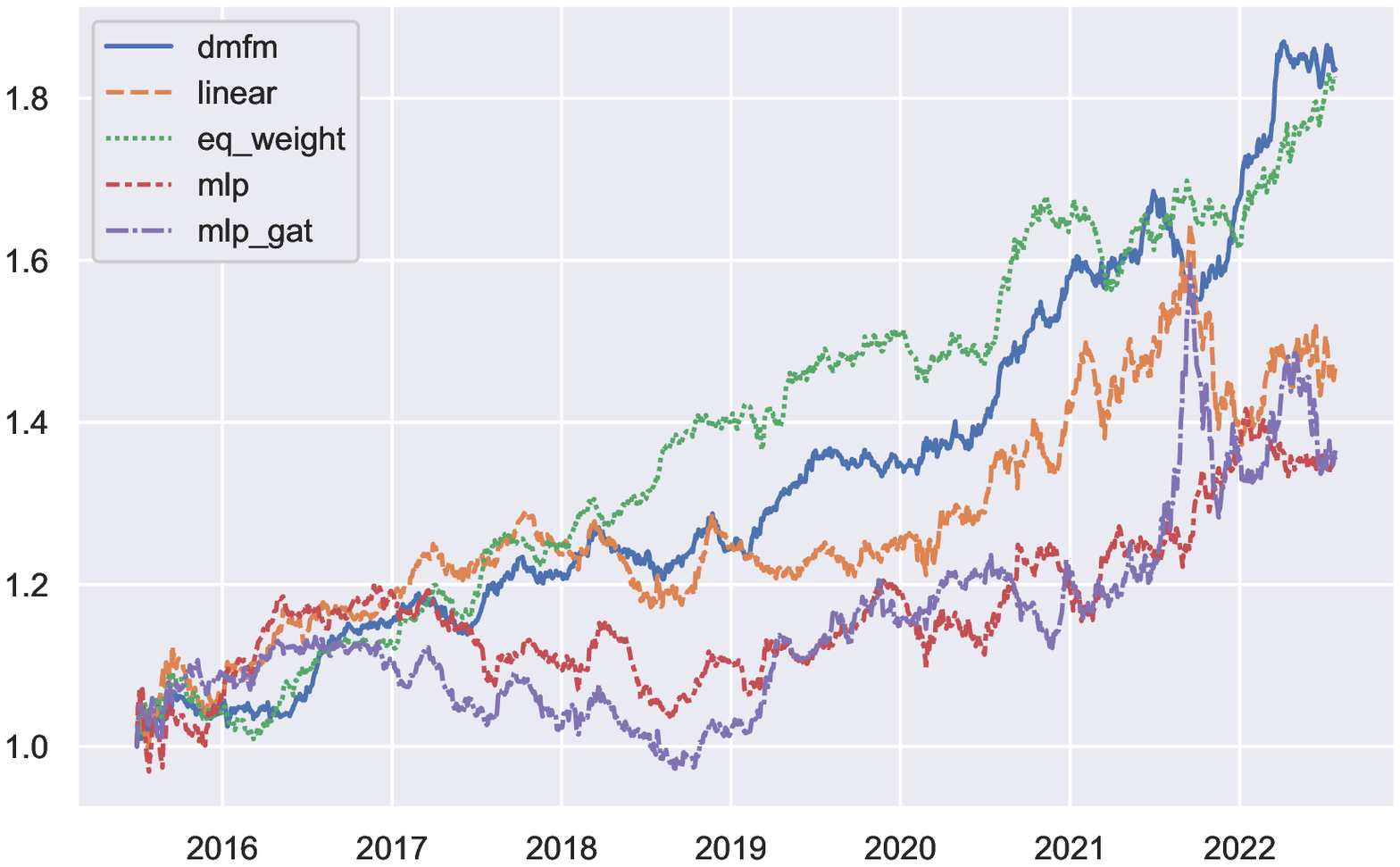}&
\includegraphics[width=0.31\textwidth]{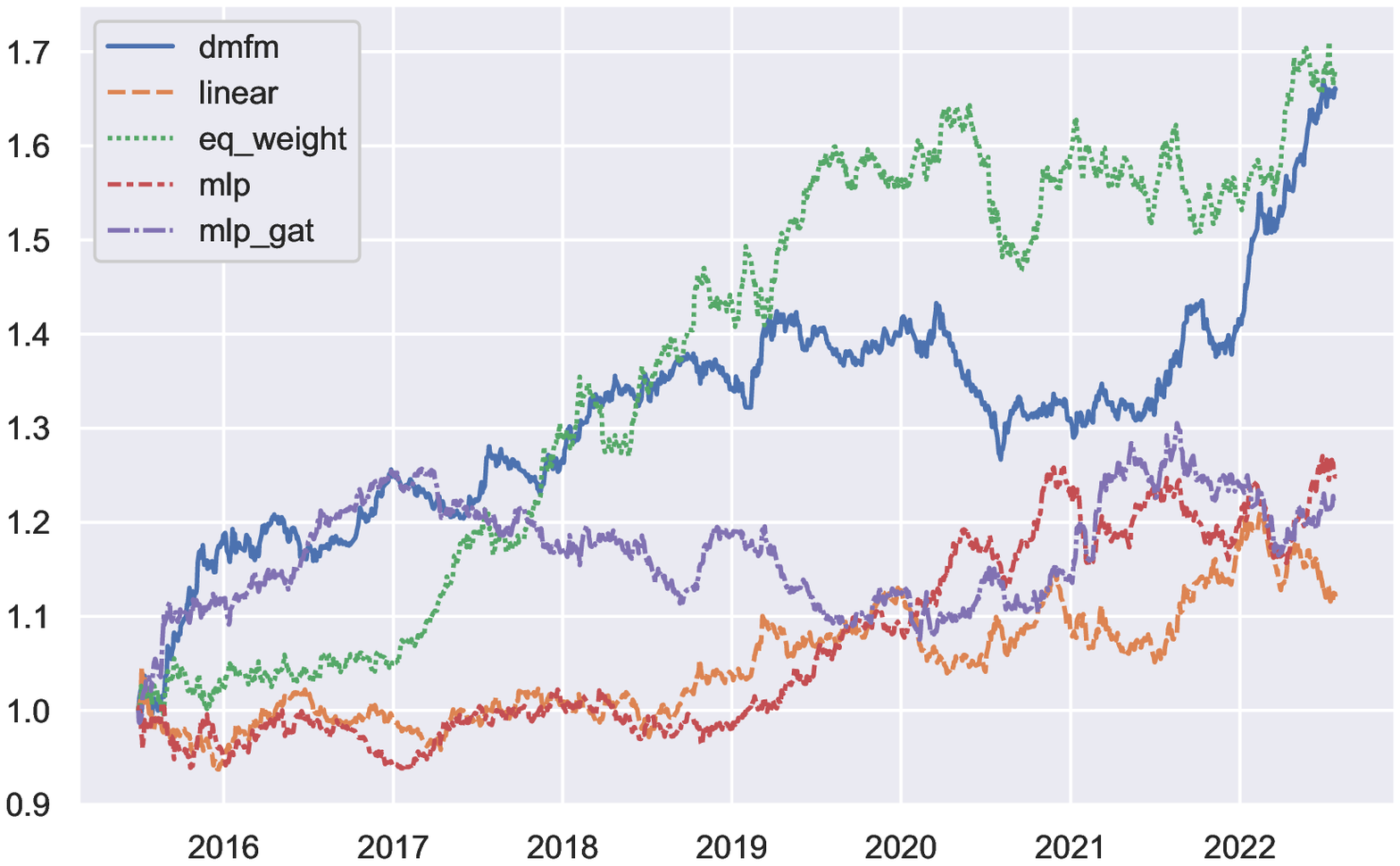}\\
\includegraphics[width=0.31\textwidth]{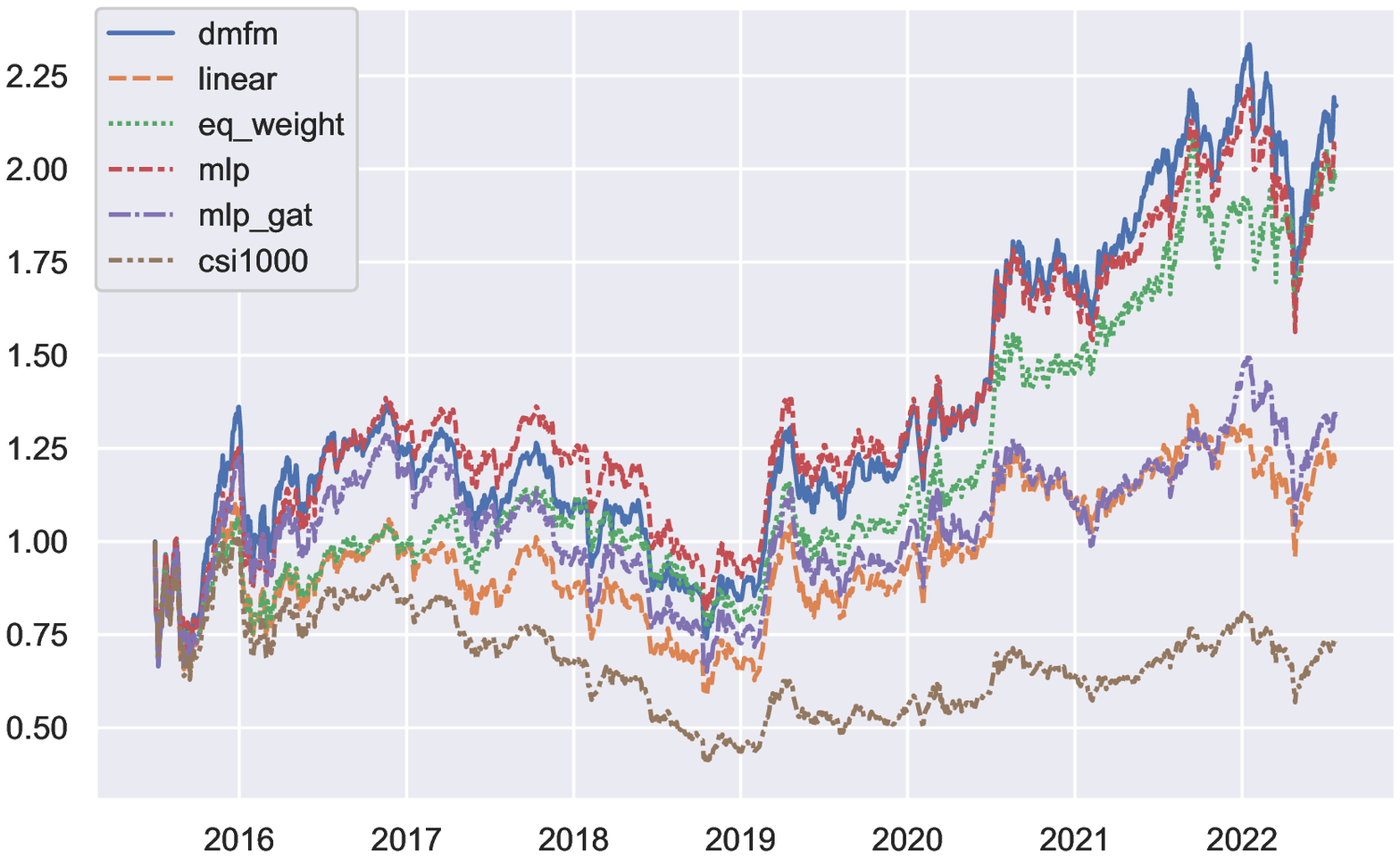}&
\includegraphics[width=0.31\textwidth]{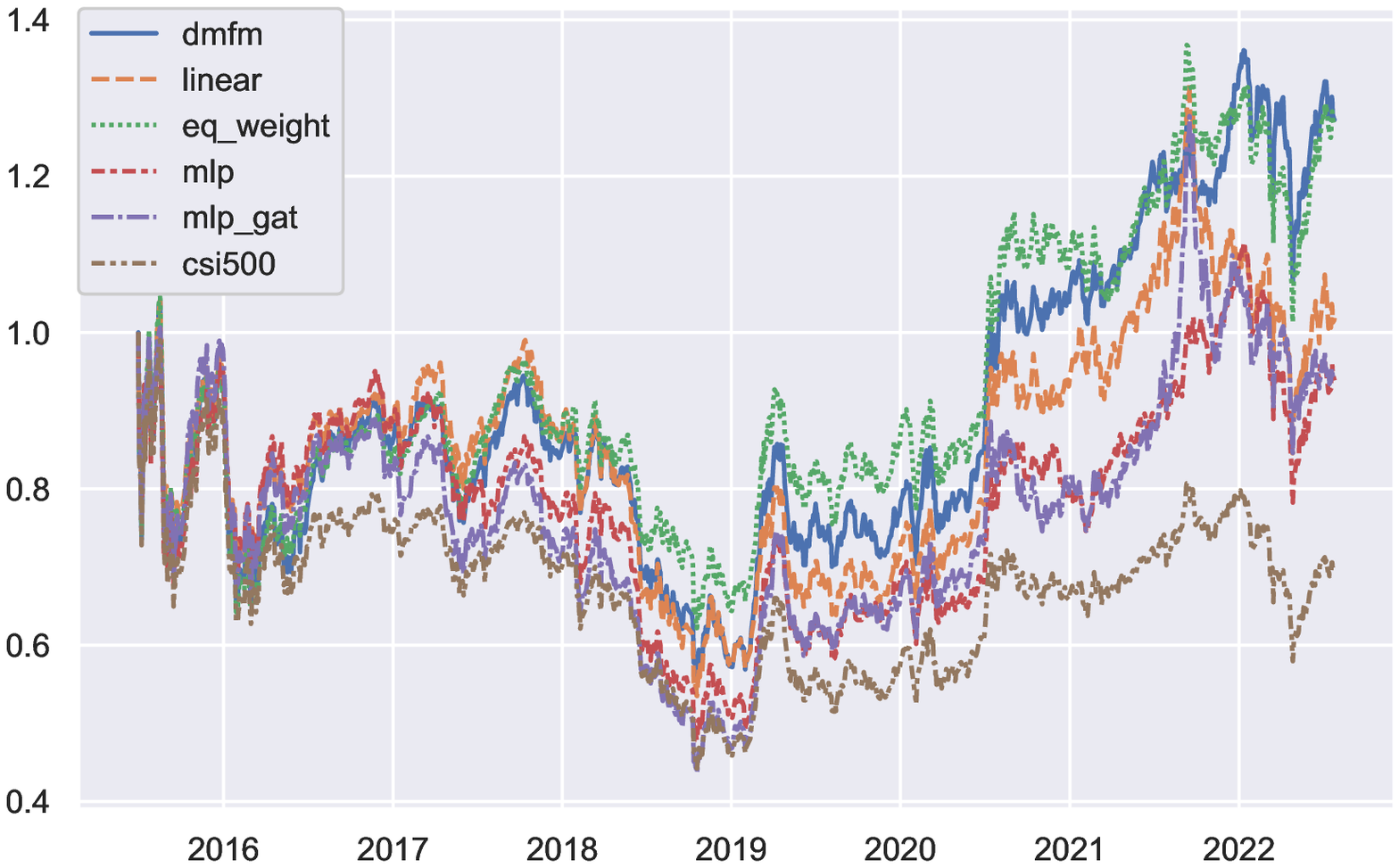}&
\includegraphics[width=0.31\textwidth]{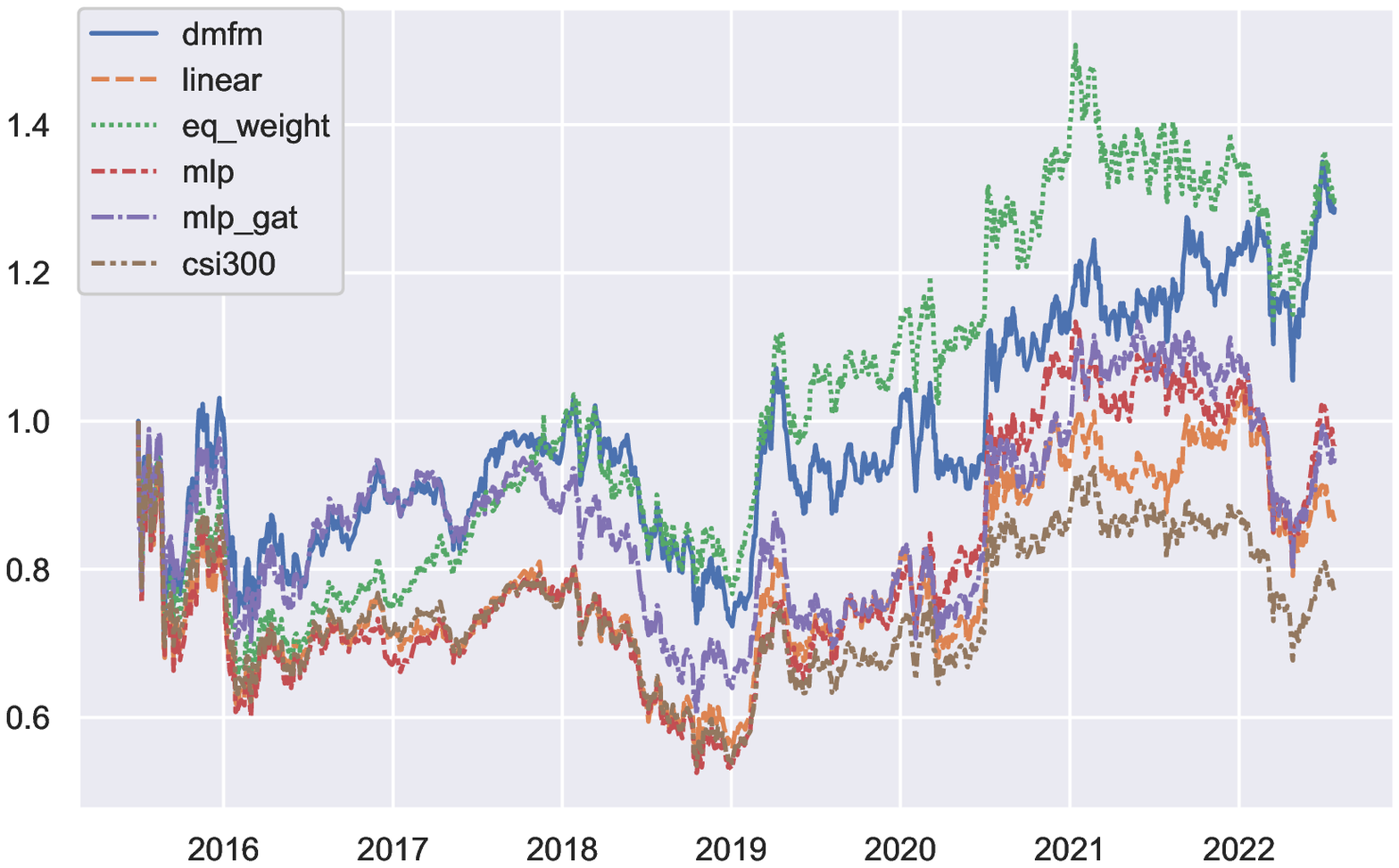}\\
CSI1000 & CSI500 & CSI300\\
\end{tabular}
\vskip -0.2cm
    \caption{Comparison of performance between our deep multifactor models and other baseline methods in different stock universes. The figures in the top rows show the curve of cumulative excess returns over each benchmark, while the second row shows the net values achieved by each method.}
    \label{fig:alpha} 
\end{figure}

\begin{table}
  \caption{Experimental results on different universes over the period 2015 to mid-2022 (\textbf{best}/\underline{2nd best})}
  \label{eval-metrics}
  \small
  \centering
  \begin{tabular}{lrrrrrrrrrrrrr}
    \toprule
    &\multicolumn{4}{c}{CSI 1000}  &\multicolumn{4}{c}{CSI 500}   &\multicolumn{4}{c}{CSI 300}                 \\
    \cmidrule(r){2-5} \cmidrule(r){6-9} \cmidrule(r){10-13}
    Method     & $\alpha$(\%)& ICIR &IR & SR&$\alpha$(\%) & ICIR &IR & SR &$\alpha$(\%) & ICIR &IR & SR\\
    \midrule
    Linear   & 7.79  & 1.08 &1.35 &0.19 &5.62 &0.89 &0.77 & 0.04 &1.73 &0.98 &0.32 &-0.09\\
    EW       & 14.65 & \textbf{1.48} &1.92& 0.47 &\underline{8.91} &\underline{1.13} &\underline{1.86} & \underline{0.18} &\underline{7.57} &0.90 &\textbf{1.20} &0.19\\
    MLP      & 15.16 & \underline{1.44} &\textbf{3.53} & \underline{0.49} &4.48 &1.03 &0.67 &-0.01&3.56 &\textbf{1.27} &0.65 &0.00\\
    MGAT     & 9.24 & 1.02 &1.63 & 0.24 &4.96 &0.83 &0.62 &0.01 &3.17 &0.48 &0.56 &-0.02\\
    Ours     & \textbf{16.36} & 1.25 &\underline{3.46} &\textbf{0.50} &\textbf{9.03} &\textbf{1.38} &\textbf{1.89} &\textbf{0.18} &\textbf{7.85} &\underline{1.13} &\underline{1.09} &\textbf{0.20}\\
    \bottomrule
  \end{tabular}
\end{table}

\begin{figure}[!htb]
   \centering
\begin{tabular}{ccc}
\includegraphics[width=0.3\textwidth]{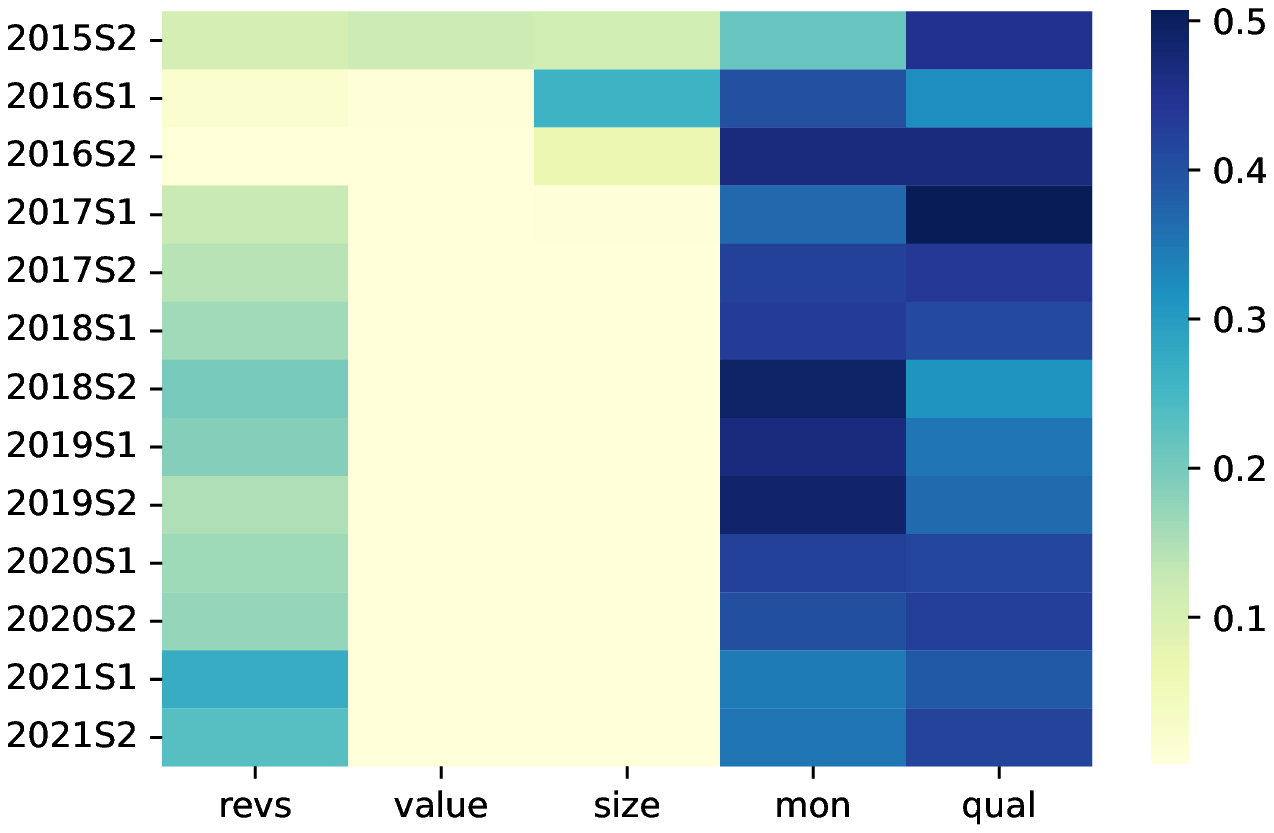}
&
\includegraphics[width=0.3\textwidth]{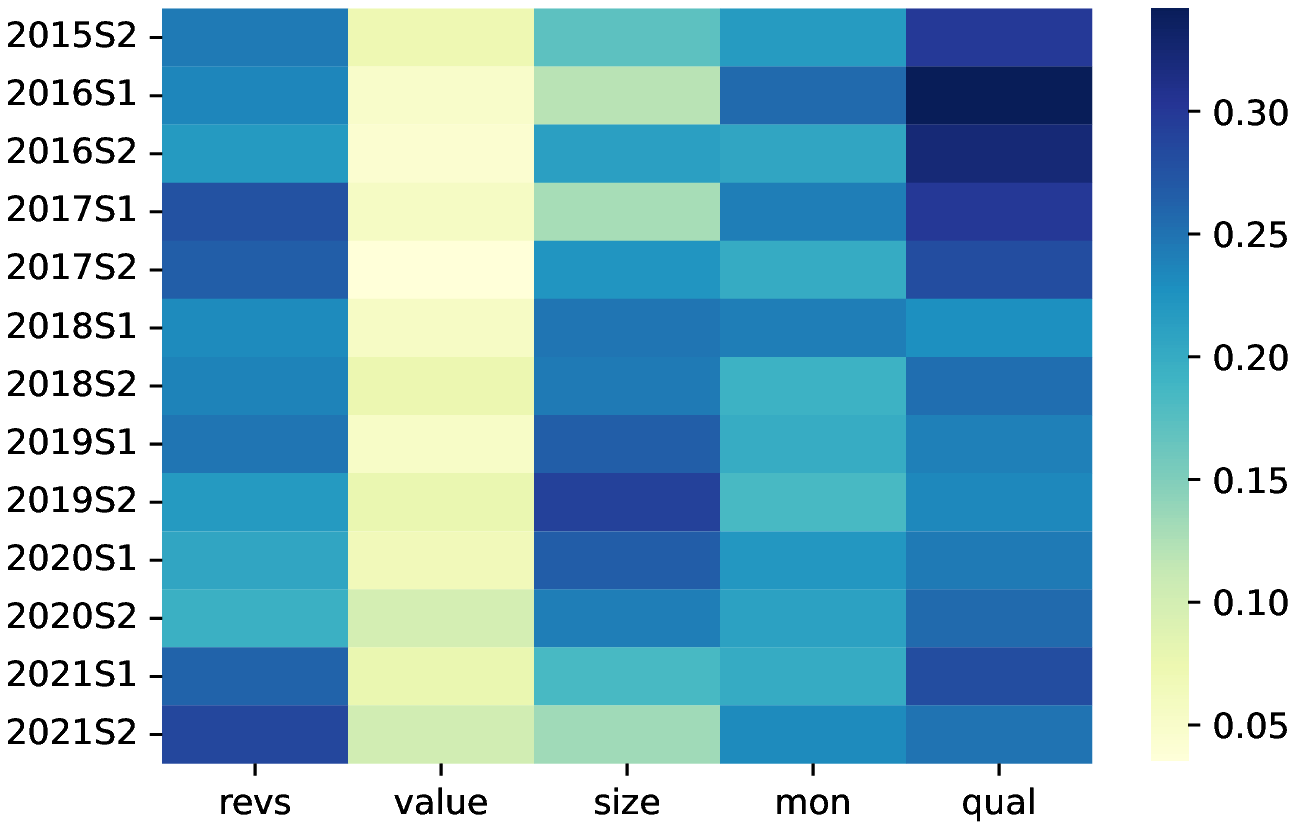}&
\includegraphics[width=0.3\textwidth]{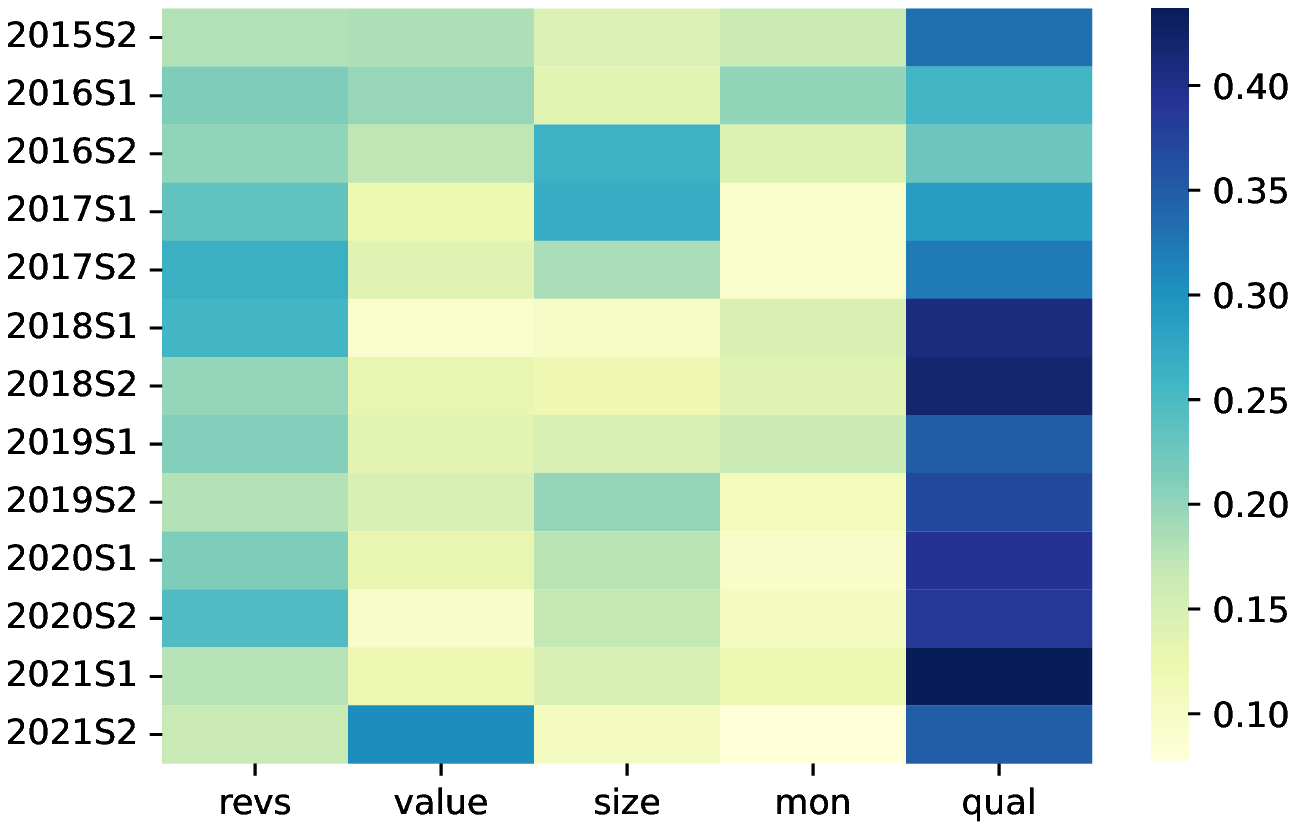}\\
\includegraphics[width=0.3\textwidth]{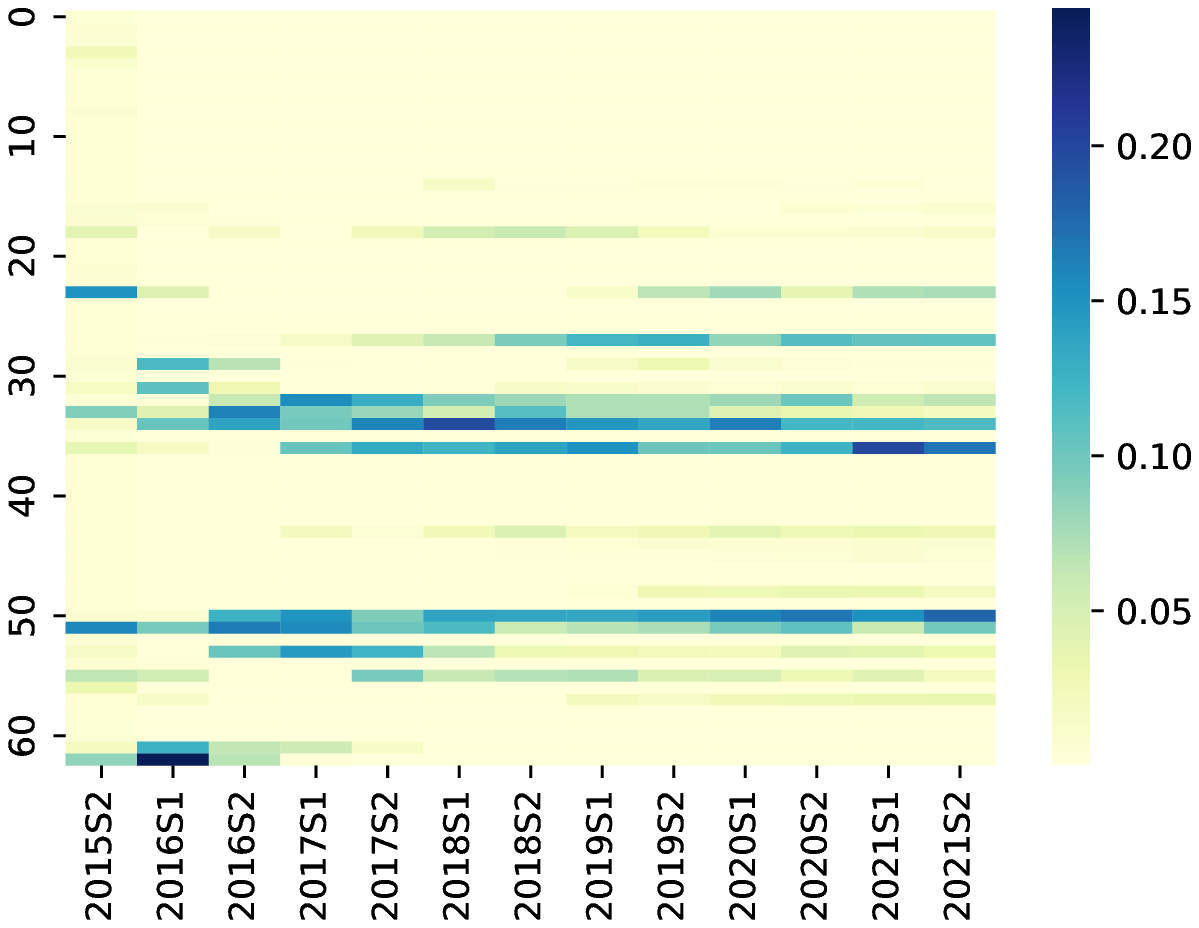}&
\includegraphics[width=0.3\textwidth]{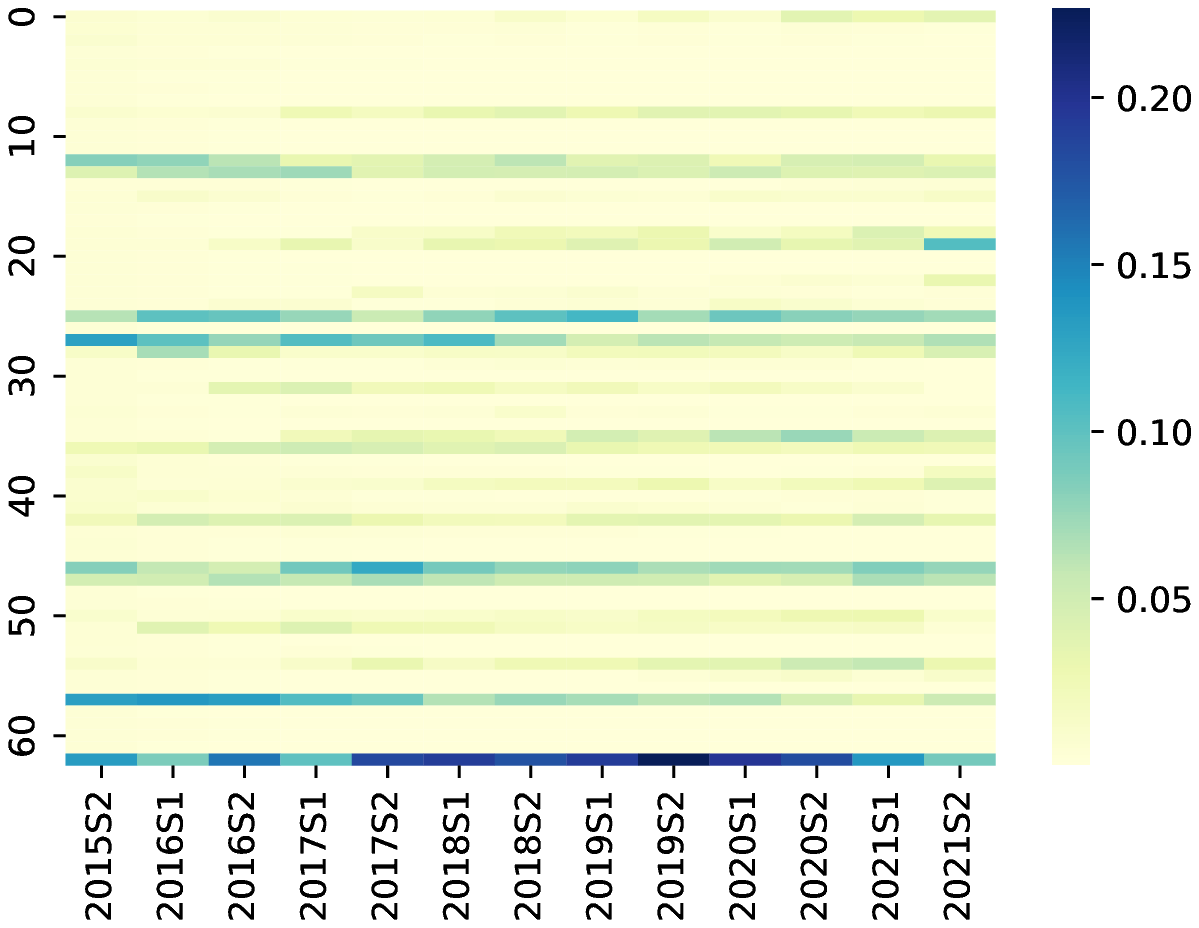}&
\includegraphics[width=0.3\textwidth]{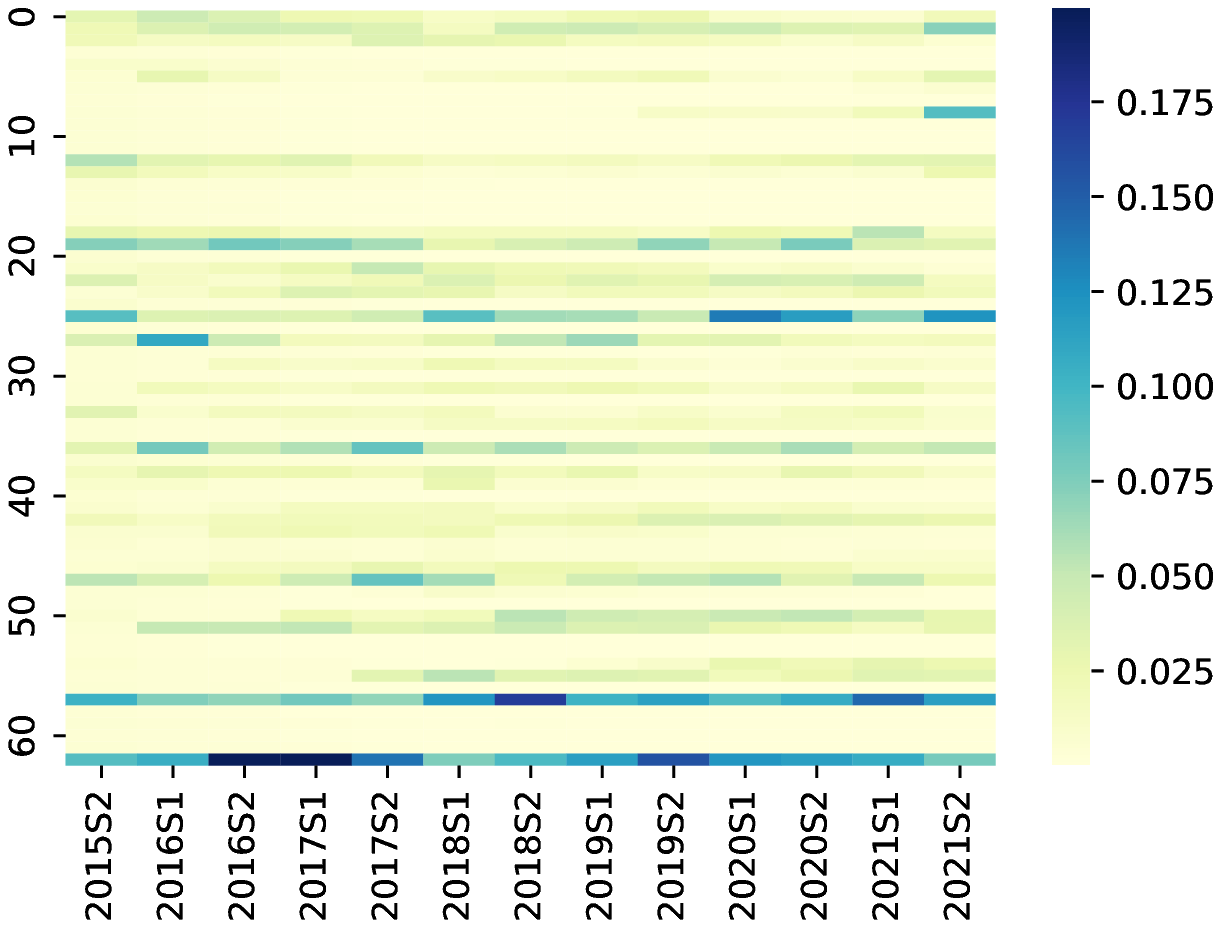}\\
CSI1000 & CSI500 & CSI300\\
\end{tabular}
\vskip -0.2cm
    \caption{The heat map shows the attention weights of the deep factors across each of the original factors: the first row shows the average attention weights for each group of factors, while the second row shows the average attention weights for each of the sixty-three style factors.}
    \label{fig:heat_map} 
\end{figure}

\begin{figure}[!htb]
\vspace{-0.4cm}
\centerline{%
\begin{tabular}{c@{}c@{}c}
\includegraphics[width=0.3\textwidth]{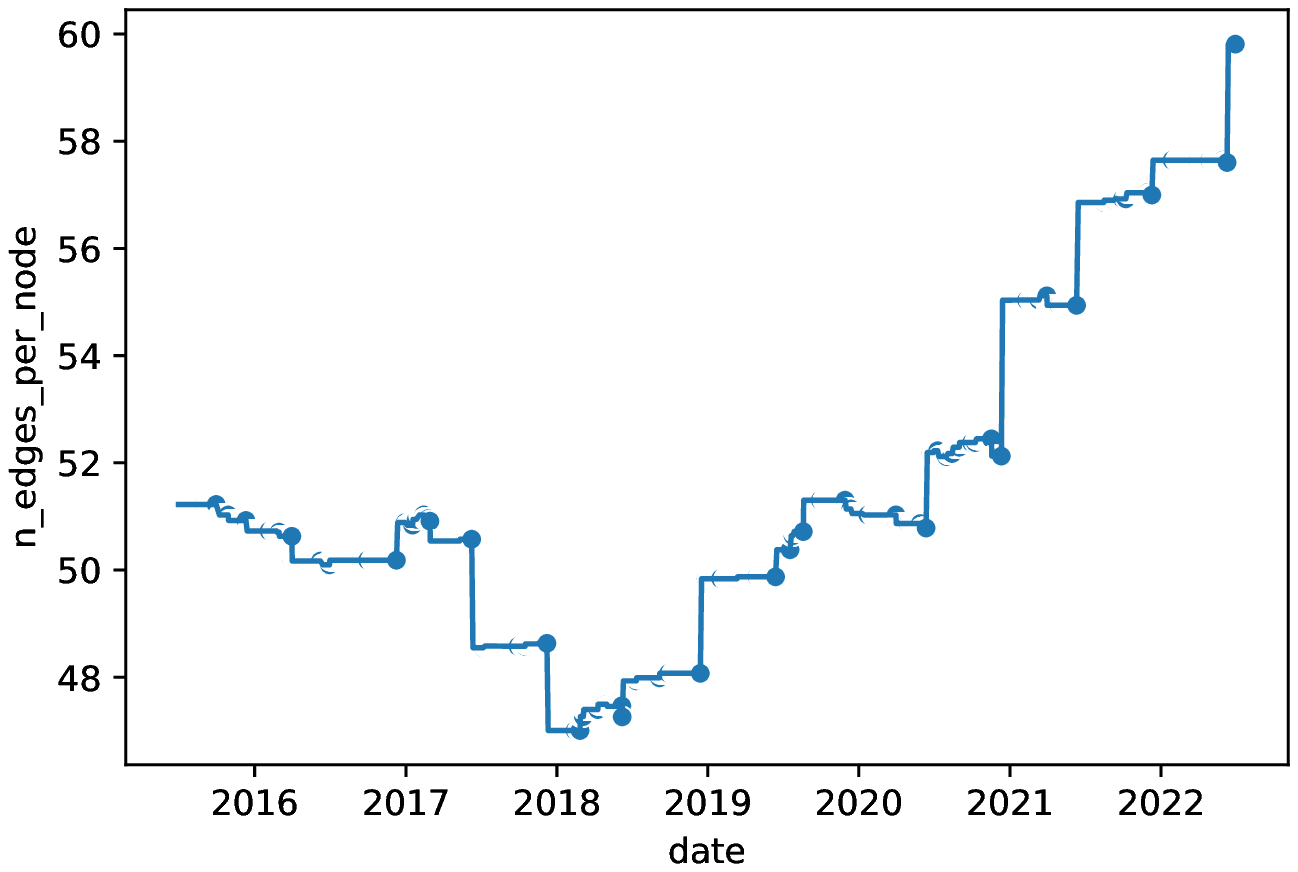} &
\includegraphics[width=0.3\textwidth]{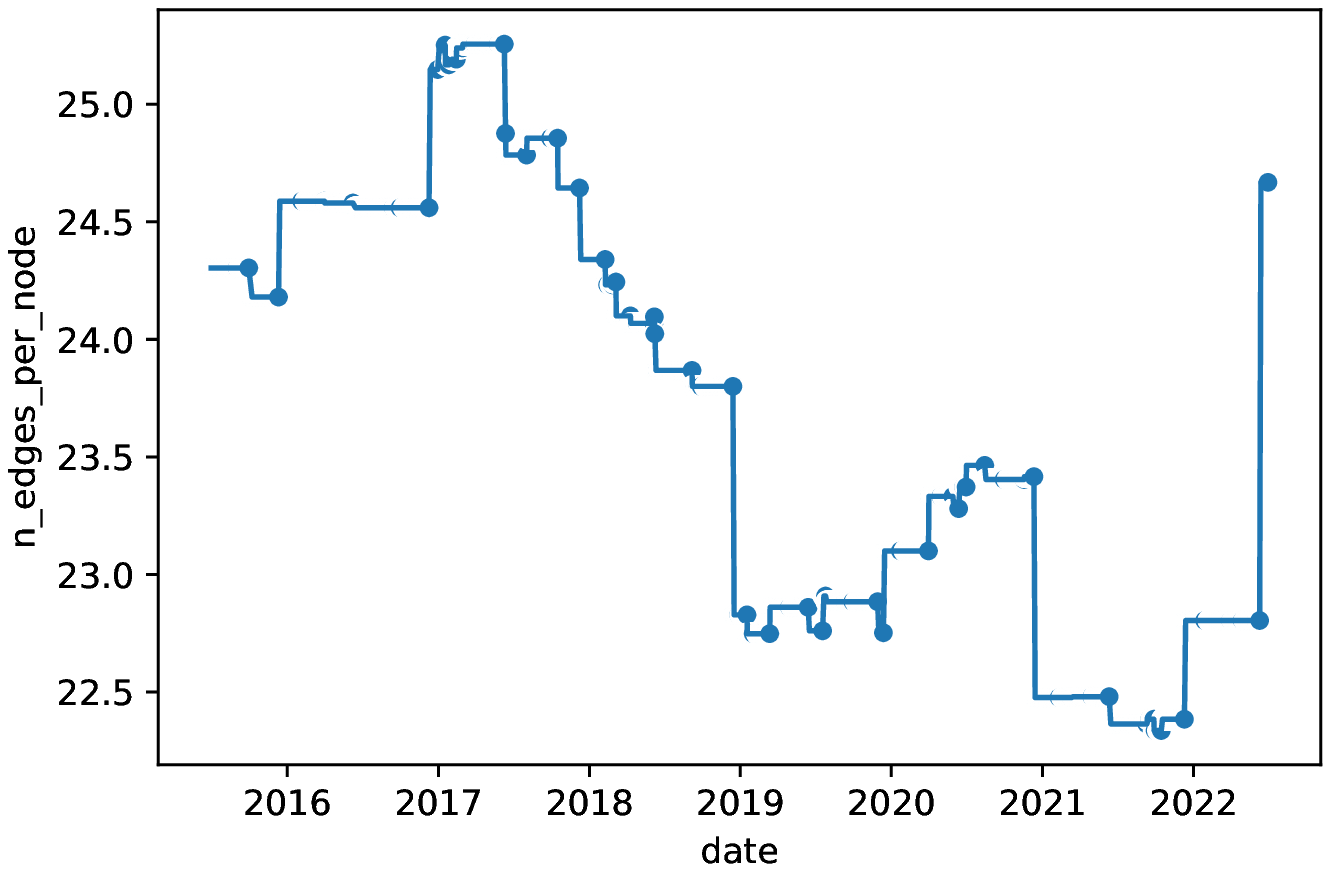} &
\includegraphics[width=0.3\textwidth]{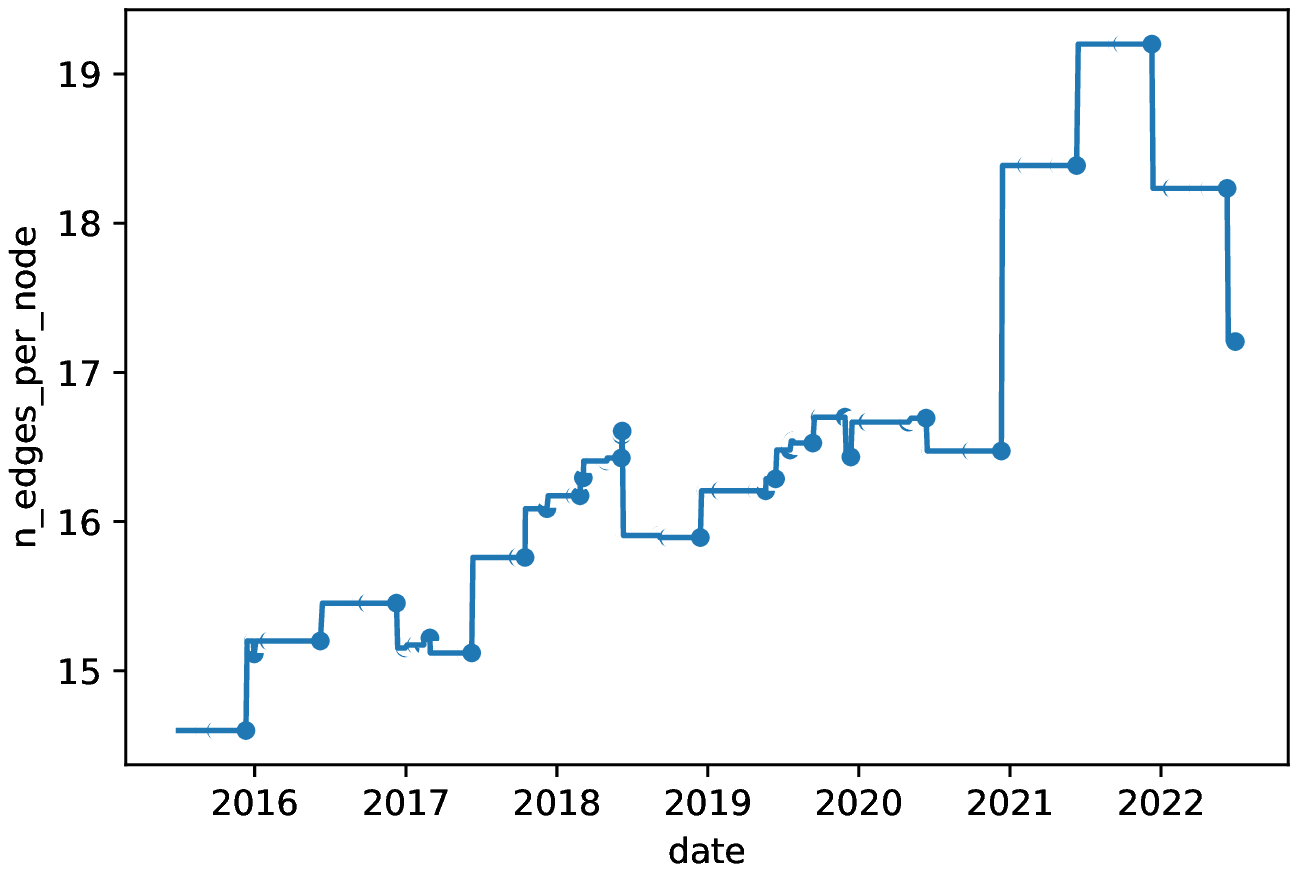} \\
(a)~~CSI1000 & (b)~~CSI500 & (c)~~CSI300
\end{tabular}}
\vskip -0.2cm
\caption{The average number of edges in the same industry (per stock)}
\label{edge}
\end{figure}

\begin{figure}[!htb]
\vspace{-0.4cm}
\centerline{%
\begin{tabular}{c@{}c@{}c}
\includegraphics[width=0.3\textwidth]{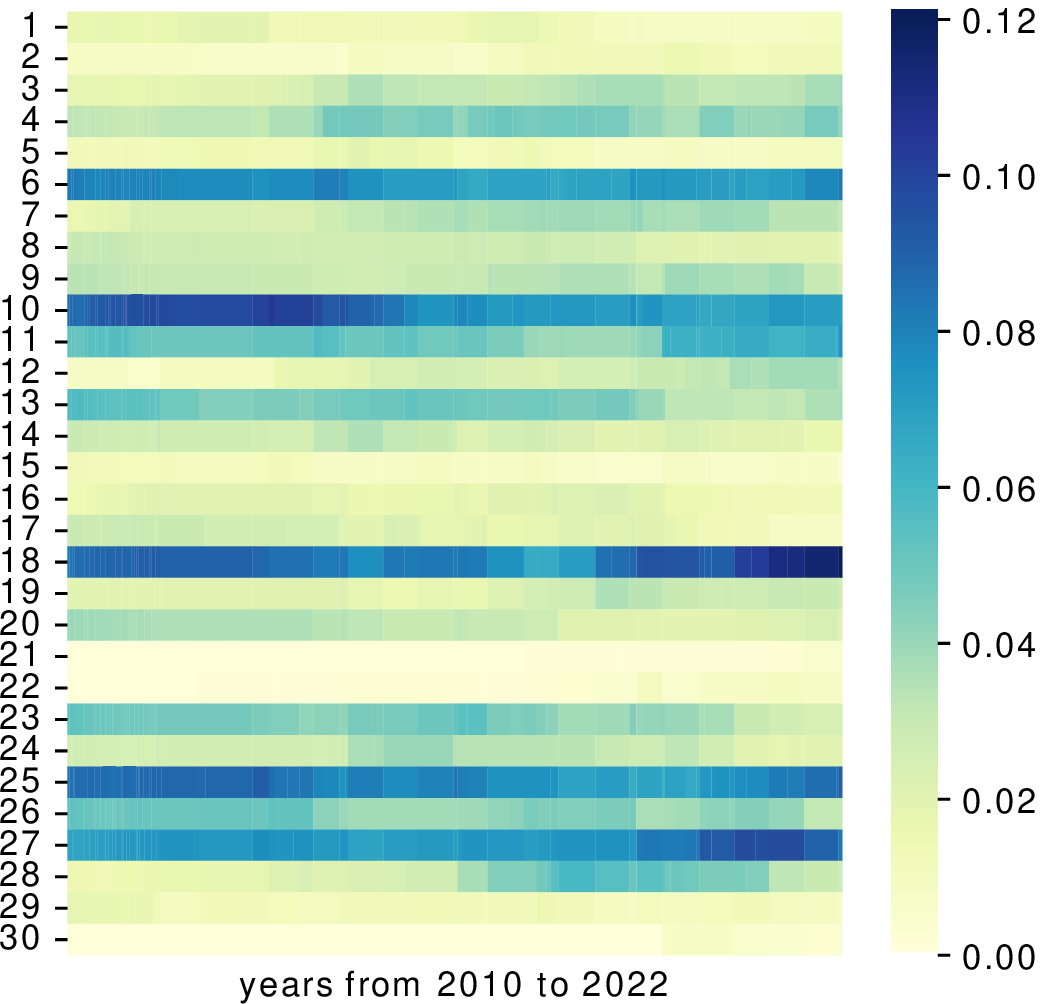} & 
\includegraphics[width=0.3\textwidth]{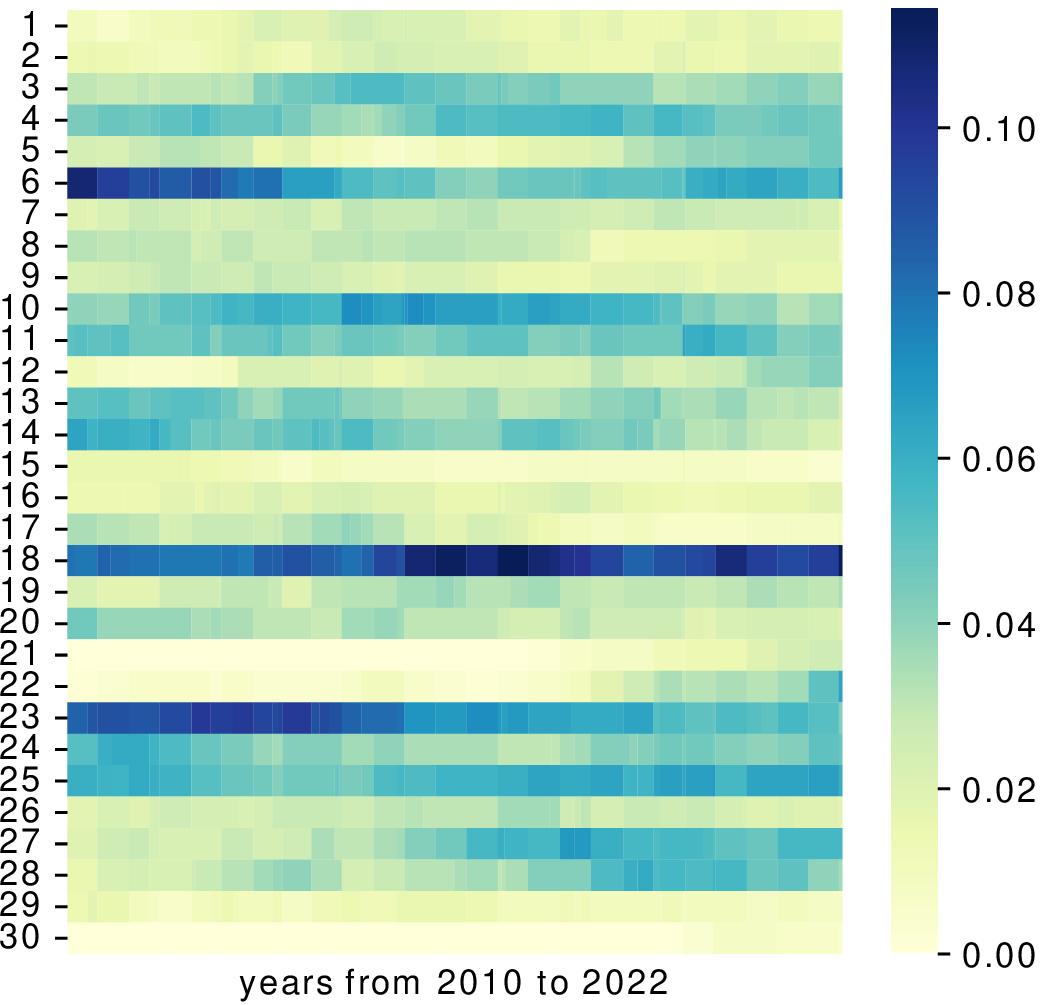} &
\includegraphics[width=0.3\textwidth]
{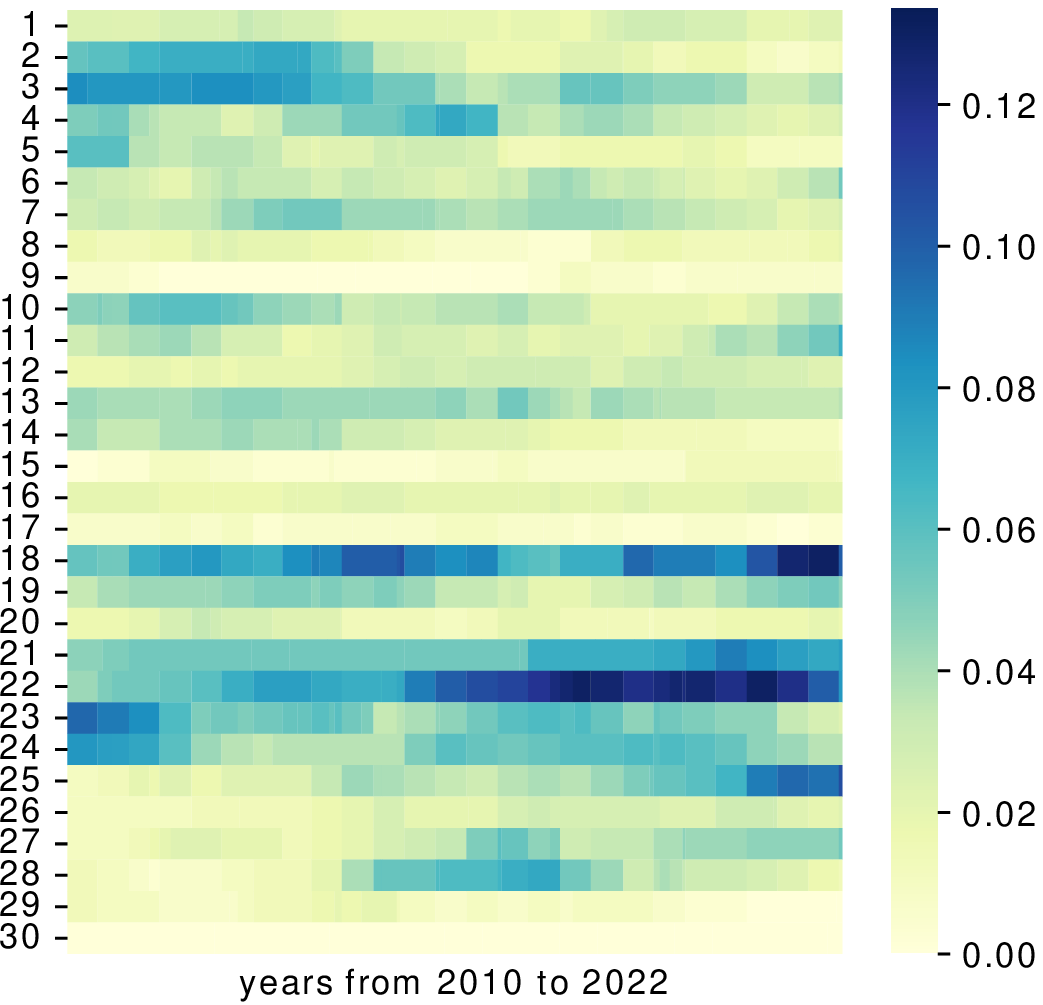} \\
(a)~~CSI1000 & (b)~~CSI500 & (c)~~CSI300
\end{tabular}}
\vskip -0.2cm
\caption{The heat map for the industry proportion in various broad-based indices from 2010 to mid-2022, with numbers from 1 to 30 representing various industries.}\label{industry}
\end{figure}

\textbf{Stock Pool.} 
In order to capture the most representative stocks of China's stock market, we select all constituents of the three major broad-based indices, namely CSI 300, CSI 500 and CSI 1000, which include 1800 stocks with the highest rank in terms of market capitalization and liquidity at the same time. Taking into account the changes caused by adjustments in the constituents, they cover more than 2800 stocks in the period from 2010 to mid-2022. By September 26, 2022, they cover 80.56\% of the total market capitalization of all stocks in the Chinese market.


\textbf {Industry Classification Standard.} For industry graph, we choose the CITIC Securities Industry Classification Standard Level 1 (CICSL1) as a guideline to define the industry relationships among stocks. The CICSL1 is widely used by institutional investors investing in Chinese market and was developed based on two well-known global standards: Global Industry Classification Standard (GICS) and Industry Classification Benchmark (ICB). Figure \ref{industry} shows the heat map of industry portion in CSI 300, CSI 500 and CSI 1000, according to CICSL1 standard.

\textbf{Dataset Construction.} Our original inputs are developed from fundamental and market data collected from publicly available data from \url{tushare.pro}. The institutional investor can also obtain the data from Refinitiv's database or WIND. We construct our dataset on a daily basis and divide the original factors into five groups derived from the Barra Global Equity Model: reversal, value, size, momentum, quality\cite{melas2018best}. 
Our dataset includes the factors that have been found to be informative in Chinese stock market \cite{lin2021learning}.
Based on the factors mentioned in \cite{melas2018best,lin2021learning}, we create other derivatives of these factors by using twelve trailing months, quarterly changes, annual changes, and factor values over yields. Finally, we create 63 factors including the base factors and their derivatives. All factors are calculated using point-in-time data, which means that there is no information leakage. For example, a company issues a financial report at time $t_1$ which is publicly available or accessible in a financial database at time $t_2$. Later, it make a restatement of that report at time $t_3$, while the information is available to investors at $t_4$. Here, $t_1$$<$$t_2$$<$$t_3$$<$$t_4$. It will never use the modified information before time $t_4$.
We follow the temporal order to divide the data into 14 data groups: Each group contains a training set, a validation set, and a test set. The training set is extended every six months. The validation set consists of the data over the following six-month period. However, we exclude the last month data from the validation set to avoid information leakage. 
The reason is that maximizing the information ratio involves using future information, \emph{i.e.}, future returns over the next $k$ trading days. Therefore, we truncate the information that might overlap with the information in the test set. 
Thus, while the test set uses data from the next six-month period, there is no overlap between the validation and test sets. For instance, the three datasets in the first group include training (2010/1/1\textasciitilde2015/6/30), validation (2015/7/1\textasciitilde2015/11/30), test (2016/1/1\textasciitilde2016/6/30), and the last group includes training (2010/1/1\textasciitilde2021/6/30), validation (2021/7/1\textasciitilde2021/11/30), test (2022/1/1\textasciitilde2022/6/30).

\textbf{Portfolio Construction.} Our goal is to select stocks from a given universe that are considered more attractive than others and build a portfolio that can outperform the benchmark of that universe. During the testing period, we select the best model with the lowest validation loss and use it to produce the deep factors. We then rank the stocks in ascending order according to their factor scores. We assume that the stocks with higher values are more attractive, which means that they could be responsible for future excess returns over the market benchmarks, \emph{i.e.}, broad-based indices. Therefore, we construct an equally weighted portfolio consisting of the 10\% most attractive stocks. The portfolio is rebalanced monthly, with transaction costs for buying and selling set at 4\textperthousand~ of the transaction amount.

\textbf{Baselines and Model Implementation.} We compare our proposed method with the following baselines: 1) \textbf{Linear} is a linear regression model where the model learns a weight for each style factor. 2) \textbf{EW} is a linear model where each descriptor (original factor) is equally weighted. EW is an important baseline that shows a result of combining factors that has nothing to do with data mining or in-sample optimization, and any deviations from equal weighting shall be determined by economic rationality \cite{melas2018best}. 3) \textbf{MLP} is a nonlinear model that learns factors from the stock context based on neural networks and consists of a context encoder and a feature decoder. 4) \textbf{MLP\& GAT } is a model based on a context encoder, a GAT, which learns the interactions between stocks in a universe graph, and a decoder, which learns a deep factor from the stock context and the universal relationships of the stocks. See supplementary materials for more details in model implementation.

\textbf{The Performance of Portfolios Constructed by Different Methods.}
In Figure \ref{fig:alpha} we show the performance curve of the portfolio created with each method, and in Table \ref{eval-metrics} we show the evaluation metrics of each method, \emph{i.e.}, the active return denoted by $\alpha$(\%), ICIR, information ratio (IR), and Sharpe ratio (SR) (see our supplementary material for more details.). These results show that our methods generally outperform others in the long run. Our deep multifactor model (DMFM) can always achieve higher active returns and Sharpe ratio than linear, EW, MLP and MLP\& GAT (MGAT) models. Although SR of the three benchmarks is -0.11, -0.20, and -0.18, respectively, our model always has a positive SR. Since CSI1000, CSI500 and CSI300 represent the small-cap, mid-cap and large-cap of Chinese stock markets, the better performance of our model in these three universes shows its generalizability at different levels of market capitalization.

\textbf{Interpretation of Deep Factor.} To make our deep factor more transparent and to clarify its insight from the financial world, we visualize the attention weight of original style factors by groups or by each of them. We take the deep factor associated with the forward twentieth trading day ($k=20$) as an example. The average attention weights over each test period are shown in Figure \ref{fig:heat_map}. In this case, the style factors belonging to the quality group receive more attention weights than other groups among all constituents of the three broad-based indexes. The heat maps for the different trading days ($k=3,5,10,15$) can be found in the supplementary materials.

\textbf{Edge Analysis for Dynamic Graph.} 
This dynamic property of a stock universe can manifest itself in changes in the relationships among stocks (see Figure \ref{edge}) and in changes in its constituents (see Figure \ref{industry}). 
A change in index components or industry classification standard (ICS) contributes to a change in the industry share for each sector in a stock universe. Since the ICS does not change too often, the change in the industry proportion is mainly caused by the change in index constituents. Therefore, Figure \ref{industry} can show most of the changes in the nodes of the stock graph. 
To illustrate the changes in the stock relationship, in Figure \ref{edge} we show the curve of average number of edges a stock has on each trading day over our industry graph in different universes (CSI1000, CSI500 and CSI300).

\section{Conclusion}
In this paper, we propose a deep cross-sectional multifactor model that incorporates financial insights and stock relationships on a dynamic and multi-relational graph in the task of factor investing. 
Our work makes three contributions: 
First, we are the first to incorporate an attention-based interpretation module into a deep learning framework that more directly shows how the original style factors explain our deep factor. 
Second, we define a stock graph with a hierarchical level of edges to represent the dynamic and multi-relational properties of stock interactions. 
Third, we build our model based on certain economic rationales and each module within our model represents a concrete financial meaning. In the future, we plan to apply our method to develop enhanced index funds (EIFs) with additional modules.


{
\small
\bibliographystyle{plain.bst}
\bibliography{dfm.bib}
}



\end{document}